%% file: main.tex
\documentclass[letterpaper, 10 pt, journal, twoside]{IEEEtran}
\usepackage{amsmath,amsfonts}
\usepackage{array}
\usepackage{textcomp}
\usepackage{stfloats}
\usepackage{url}
\usepackage{verbatim}
\usepackage{graphicx}
\usepackage{cite}
\usepackage{mathtools}
\usepackage{hyperref}
\usepackage{color}
\usepackage[dvipsnames]{xcolor}
\usepackage{tabularx}
\usepackage{caption} 

\input{./content/macros}

\usepackage{mdframed}
\mdfdefinestyle{MyFrame}{%
    linecolor=black,
    outerlinewidth=2pt,
    innertopmargin=4pt,
    innerbottommargin=4pt,
    innerrightmargin=4pt,
    innerleftmargin=4pt,
    leftmargin = 4pt,
    rightmargin = 4pt,
    skipabove=10pt,  % Space above the box
    skipbelow=10pt  % Space below the box
        }

\begin{document}

\title{DualGuard MPPI: Safe and Performant Optimal Control by Combining Sampling-Based MPC and Hamilton-Jacobi Reachability}

% Make room for more info lines in the \author command 
\author{Javier Borquez$^{1}$, Luke Raus$^{2}$, Yusuf Umut Ciftci$^{1}$, and Somil Bansal$^{3}$ %
%Use only for final RAL version
\thanks{This work is supported by the University of Santiago de Chile, the NSF CAREER Program under award 2240163 and the DARPA ANSR program.} %Use only for final RAL version
\thanks{$^{1}$Javier Borquez and Yusuf Umut Ciftci are with University of Southern California.
        {\tt\footnotesize javierbo@usc.edu}}%
\thanks{$^{2} $Luke Raus is with Olin College of Engineering.}
\thanks{$^{3} $Somil Bansal is with Stanford University. {\tt\footnotesize somil@stanford.edu}}
}
% Use only for final RAL version.

\markboth{}
{Borquez \MakeLowercase{\textit{et al.}}: DualGuard MPPI - Safe Optimal Control via Sampling MPC and HJ Reachability} 

\colorlet{usercolorname}{red!00} %make nonzero to highlight
\sethlcolor{usercolorname}
%%%%%%%%%%%%%%%%%%%%%%%%%%%%%%%

\setcounter{figure}{-2}
\makeatletter
\let\@oldmaketitle\@maketitle
    \renewcommand{\@maketitle}{\@oldmaketitle
    \centering
    \vspace{1em}
    \includegraphics[width=0.99\textwidth]{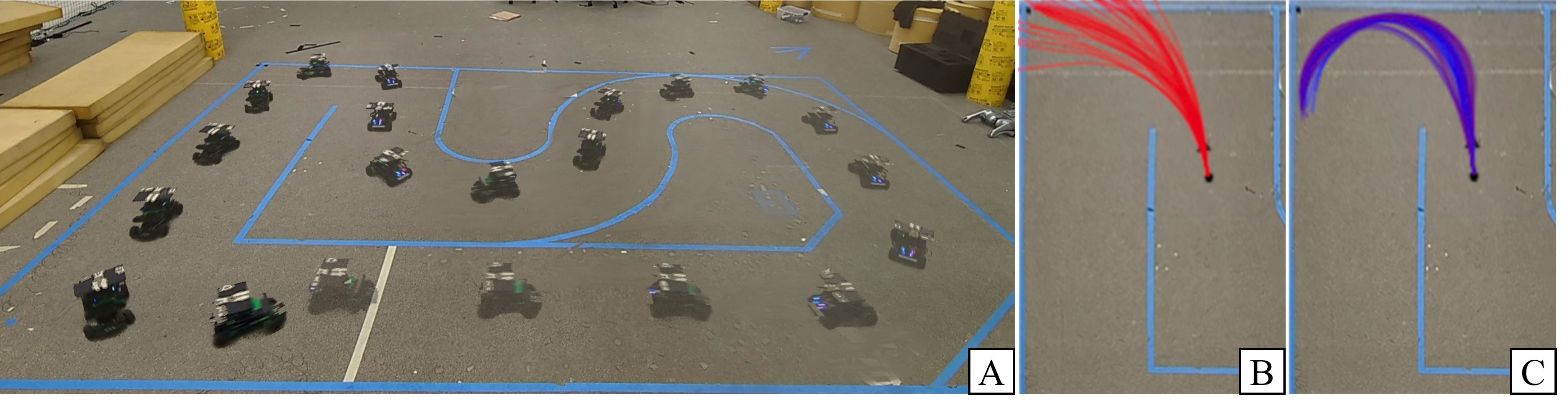}
    \vspace{-0.4cm}
    
    \captionof{figure}{We propose DualGuard MPPI - a framework to solve optimal control problems subjected to hard safety constraints. Our method integrates safety filtering during the sampling process in MPPI to generate safe rollouts, ensuring safe executions while improving sample efficiency. An output least restrictive filter is used to ensure safe executions on the system, despite potential multimodality in the sampling process. (A) We apply the proposed framework to an RC car experiment where the vehicle completes laps without leaving the track (breaching safety), while trying to stay centered in the lane and going as fast as possible. (B) Sampled trajectories in classical MPPI generate only high-cost unsafe executions near a tight corner, which results in breaching the boundary of the track. (C) DualGuard MPPI safe rollouts generate only collision-free executions with mild costs depending only on performance criteria, resulting in safe and performant behavior. Details on this experiment are provided in Section \ref{case_rc_car}.}
    \vspace{-0.7cm} % this command is important for spacing!!! dont comment or delete it -- zc after long tuning.
    \label{fig:exp_main_result}
    
  % \bigskip
  }
\makeatother
%%%%%%%%%%%%%%%%%%%%%%%%%%%%%%%
\maketitle

\begin{abstract}
Designing controllers that are both safe and performant is inherently challenging.
This co-optimization can be formulated as a constrained optimal control problem, where the cost function represents the performance criterion and safety is specified as a constraint.
While sampling-based methods, such as Model Predictive Path Integral (MPPI) control, have shown great promise in tackling complex optimal control problems, they often struggle to enforce safety constraints. 
To address this limitation, we propose DualGuard-MPPI, a novel framework for solving safety-constrained optimal control problems.
Our approach integrates Hamilton-Jacobi reachability analysis within the MPPI sampling process to ensure that all generated samples are provably safe for the system. 
On the one hand, this integration allows DualGuard-MPPI to enforce strict safety constraints; at the same time, it facilitates a more effective exploration of the environment with the same number of samples, reducing the effective sampling variance and leading to better performance optimization.
Through several simulations and hardware experiments, we demonstrate that the proposed approach achieves much higher performance compared to existing MPPI methods, without compromising safety.
\end{abstract}

%\vspace{-0.5cm}
% 
\input{./content/01_introduction.tex}
\input{./content/02_problem.tex}
\input{./content/03_background.tex}

\input{./content/04_approach.tex}

% 
\input{./content/05_results.tex}
%
\input{./content/06_hardware}

\input{./content/07_conclusion.tex}
%
\addtolength{\textheight}{-1cm}   % This command serves to balance the column lengths
% on the last page of the document manually. It shortens         % the textheight of the last page by a suitable amount.
\bibliographystyle{IEEEtran}
\bibliography{./bib_local}

\end{document}

%% file: content/macros.tex
\usepackage{graphicx,caption,subcaption}
\usepackage{amsmath,amssymb,stmaryrd,mathtools}
\usepackage[ruled,vlined,titlenotnumbered,linesnumbered]{algorithm2e} 
\usepackage[noend]{algpseudocode}
\usepackage{acronym}
\usepackage{comment}
\usepackage{color}
\usepackage{units}
\usepackage{color, colortbl}
\usepackage{adjustbox}
\usepackage{array,multirow}
\usepackage{xcolor}
\usepackage{soul}
\usepackage{wrapfig}

\definecolor{darkGray}{gray}{0.7}
\definecolor{lightGray}{gray}{0.9}

\usepackage[capitalise]{cleveref}

\DeclareCaptionLabelSeparator{periodspace}{.\quad}
\crefname{equation}{}{} % Default to eqref for equations
\crefname{section}{Sec.}{Sec.}

\newcommand{\tvar}{t}
\newcommand{\thor}{T} % Time horizon. could be t_f

\newcommand{\tinit}{0}
\newcommand{\tfinal}{t_{f}}

\newcommand{\ctrl}{u}
\newcommand{\dstb}{d}
\newcommand{\cfunc}{u(\cdot)}
\newcommand{\dfunc}{d(\cdot)}
\newcommand{\cset}{\mathcal{U}}

\newcommand{\dset}{\mathcal{D}}

\newcommand{\state}{\mathbf{x}}
\newcommand{\sset}{\mathcal{X}}
\newcommand{\traj}{\xi} 
\newcommand{\dyn}{f} 
\newcommand{\targetfunc}{l}
\newcommand{\targetset}{\mathcal{F}}
\newcommand{\costfunctional}{J}

\newcommand{\vfunc}{V}
\newcommand{\brs}{\mathcal{V}} % backwards reachable set

\newcommand{\ham}{H}

\newcommand{\ctrlseq}{\mathbf{\ctrl}}
\newcommand{\dstbseq}{\mathbf{\dstb}}

\newcommand{\tdummy}{\tau}

\newcommand{\controller}{u}
\newcommand{\ctrlnom}{u_{\text{nom}}}
\newcommand{\strat}{\gamma}
\newcommand{\stratset}{\Gamma}

\newcommand{\horizon}{T}

\newcommand{\argmax}{\operatornamewithlimits{argmax}}
\newcommand{\argmin}{\operatornamewithlimits{argmin}}

%% file: content/01_introduction.tex
\section{Introduction}

\IEEEPARstart{C}{o-optimizing} safety and performance is a critical challenge in the design of controllers for autonomous systems, especially in safety-critical domains such as autonomous vehicles, UAVs, and robotics. In such settings, controllers must ensure that performance objectives are met while guaranteeing that safety constraints are respected. This can be framed as a constrained optimal control problem, where the goal is to satisfy safety constraints throughout the trajectory while optimizing performance objectives.

Several methods have been proposed to address this co-optimization challenge. Dynamic programming (DP) offers a rigorous solution to constrained optimization \cite{one_stage_DP,Wang2024coopt_constrained}. Still, it is computationally intractable for many real-time applications due to the ``curse of dimensionality''. %, often associated with DP-based solutions
On the other hand, MPC is more feasible for real-time applications, leveraging its ability to generate optimized control sequences over a receding horizon \cite{schwenzer2021review,borrelli2017predictive}. 
Among these, Model Predictive Path Integral (MPPI) control is a sampling-based MPC method that has gained popularity due to its flexibility in handling complex dynamics, uncertainties, and non-linear systems \cite{mppi_tro}. 
MPPI leverages stochastic sampling to optimize control trajectories, offering a scalable and efficient way to generate control actions. 
However, ensuring safety within MPPI has proven challenging, as the basic framework does not account for hard safety constraints. 
Consequently, various extensions have been proposed to incorporate safety, each solving the co-optimization problem differently.

Classical MPPI-based approaches encourage safety by penalizing unsafe sampled trajectories, which are then mostly ignored in the sample aggregation process.
However, this approach cannot guarantee safety and its satisfaction depends on the penalty function. Another drawback is that computation is wasted on ignored samples, reducing the effective number of samples used for optimization.
To enforce safety, approaches such as safety filtering and barrier functions have emerged, where constraints are enforced as a post-optimization correction \cite{borquezFiltering2023,ames2016control}. 
Such approaches allow to solve the MPPI problem first and apply safety filtering afterward, or combine these two approaches \cite{mppi_theory,mppi_shield,wabersich2021predictive}.
However, these approaches tend to ignore the long-term effect of safety actions on performance. 
Other procedures use probabilistic methods to address the safety-performance co-optimization, such as modifying the sampling distribution to favor safer trajectories \cite{safe_importance_sampling}, using stochastic safety certificates to steer away from unsafe regions \cite{learn_stochastic_barrier}, or improving safety indirectly by enhancing adaptability and robustness to uncertainty \cite{proactive_mppi, unscented_mppi}.

Beyond MPPI literature, control barrier function methods, allow for the co-optimization of safety and performance by solving constrained optimization problems \cite{ames2016control, CBF_MPC}. However, these approaches tend to be myopic, optimizing safety and performance locally without capturing long-term effects, and often rely on Lyapunov-based objectives that limit the expressiveness of performance goals. Filter-aware motion planning has also been explored to mitigate this performance degradation caused by safety interventions, where instead of treating safety reactively, these approaches explicitly account for future interventions. For instance, shielding-aware planning proactively balances performance with the risk of emergency maneuvers triggered by low-probability agent behaviors \cite{SHARP, SHARP2}. Similarly, in reinforcement learning settings, integrating safety filters during training rather than solely during execution has been shown to improve performance and reduce undesired safety interventions\cite{reduce_filter, Filter_train_RL}.

In this work, we introduce DualGuard-MPPI, a novel MPPI algorithm that incorporates safety constraints directly into the MPPI algorithm using Hamilton-Jacobi (HJ) reachability. 
Our key idea is to incorporate two safety filtering stages in the MPPI algorithm: first, we perform safety filtering along the sampled control perturbations, leading to provably safe rollouts of the system trajectories, which are then used to optimize the performance objective. 
Second, we apply safety filtering on the selected control sequence before it is applied to the system to prevent multimodal combinations that may compromise safety. In this context, multimodality refers to distinct groups of safe trajectories, where different evasive maneuvers—such as swerving left or right to avoid an obstacle—can both be viable but should not be arbitrarily mixed.

The proposed approach imposes safety constraints throughout the trajectory optimization and control application stages, ensuring that the system operates within safe bounds at all times, without requiring any safety penalty tuning.
Moreover, since all generated samples are provably safe, they all contribute to performance optimization, effectively reducing the MPPI sampling variance and leading to a better performance, as compared to vanilla MPPI algorithm.
In summary, the key contributions of this work are: 

\begin{itemize}
  \item An MPPI framework that co-optimizes safety and performance while ensuring safety at all times.
  \item Elimination of safety penalty tuning within MPPI and improvement in sampling efficiency by guaranteeing that all sampled control sequences are safe.
  % \item Performance-only optimal control problem formulation, as safety is guaranteed in the sampling stage.
  \item Extensive simulations and hardware experiments demonstrating the superior performance and robustness of the proposed approach without compromising safety.
  % \item Providing safety guarantees in the presence of combinations of individually-safe multimodal executions.
\end{itemize}

%% file: content/02_problem.tex
\section{Problem Statement}

Consider an autonomous system with state $\state \in \sset \subseteq \mathbb{R}^n$ that evolves according to dynamics $\dot{\state} = \dyn(\state, \ctrl, \dstb)$, where $\ctrl \in \cset$ and $\dstb \in \dset$ are the control and disturbance of the system, respectively. 
$\dstb$ can represent potential model uncertainties or an actual, adversarial exogenous input to the system.
We assume the dynamics are uniformly continuous in $u$ and $d$, bounded, and Lipschitz continuous in $\state$ for fixed $\ctrl$ and $\dstb$.
Finally, let $\traj_{\state,\tvar}^{\ctrlseq,\mathbf{\dstb}}(\tdummy)$ denote the system state at time $\tdummy$, starting from the state $\state$ at time $\tvar$ under control signal $\ctrlseq(\cdot)$ and disturbance signal $\dstbseq(\cdot)$ while following the dynamics. A control signal $\ctrlseq(\cdot)$ is defined as a measurable function mapping from the time horizon to the set of admissible controls $\cset$, and a disturbance signal is similarly defined. 
We additionally assume that the control and disturbance signals $\ctrlseq(\cdot)$ and $\dstbseq(\cdot)$ are piecewise continuous in $t$. 
This assumption ensures that the system trajectory $\traj_{\state,\tvar}^{\ctrlseq,\mathbf{\dstb}}$
exists and is unique and continuous for all initial states \cite{coddington1955theory, callier2012linear}.

In addition, we are given a failure set $\targetset \subset X$ that the system must avoid at all times (e.g., obstacles for a navigation robot). The safety constraint is encoded via a Lipschitz continuous function $
\targetfunc (\cdot): \targetset= \{\state : \targetfunc(\state)\leq 0\}$.
We aim to design a controller that optimally balances the system's safety constraints and performance objectives. The performance objective is given by minimizing a cost function $S$ over the system trajectory, given by:
\begin{equation}    S(\traj)=\phi(\state(\tfinal),\tfinal)+\int_{\tinit}^{\tfinal}{\mathcal{L}}(\state(\tvar),\ctrl(\tvar),\tvar)\mathrm{d}\tvar
\end{equation}
where $\phi$ and $\mathcal{L}$ represent the terminal and running cost respectively, and $\tfinal$ the task completion time.
Specifically, we aim to ensure that the control actions $\ctrl$ drive the system towards minimizing $S$ while rigorously maintaining system safety by preventing the state $\state$ from entering $\targetset$ even for the worst case disturbances $\dstb$. This takes the form of the following constrained optimal control problem:
\begin{equation}\label{eq:opt_problem}
\begin{aligned}
    \ctrl^{*}(\cdot)&=\argmin_{\ctrl(\cdot)}S(\traj_{\state,\tvar}^{\ctrlseq,\mathbf{\dstb}},\ctrl(\cdot)) \\ 
    \text{s.t.}\quad
    \dot{\state}(t) &= \dyn(\state(t), \ctrl(t), \dstb(t)), \\
    \quad \;\targetfunc(\state(\tvar))& > 0, 
    \;\ctrl(t) \in \cset, 
    \;\dstb(t) \in \dset,
   \forall t \in [\tvar, \tvar_f]
\end{aligned}
\end{equation}
\noindent the optimization problem defined in \eqref{eq:opt_problem} in general is non-convex and can be difficult to solve. In this work, we propose a novel MPPI method to solve this problem.

%% file: content/03_background.tex
\section{\label{background}Background}

\subsection{\label{background_mppi}Model Predictive Path Integral (MPPI)}

Original MPPI~\cite{mppi_tro} aims to solve a stochastic optimal control problem using a zeroth-order sampling-based MPC scheme for undisturbed deterministic general nonlinear systems.
Specifically, at each state, MPPI samples K random control sequences around a nominal control sequence for a time horizon $\hat{T} \leq \tvar_f$ discretized over $H$ evenly spaced steps. The control input for the \textit{k}th sequence at \textit{j}th time step is given by: $u^k(x, j) = u(x, j) + \delta_j^k$, where $u(x, j)$ is a nominal control sequence and $\delta_j^k$ a randomly sampled control perturbation. Next, the state trajectory $\xi_j^k$ and the cost to go $S(\xi_j^k)$ corresponding to each control sequence is computed. It is shown that the optimal control sequence can be approximated by the following update law~\cite{mppi_theory}:
\begin{equation}\label{eq:update_law}
u(\state,j)^{\ast}\approx u(\state,j)+{\frac{\sum_{k=1}^{K}\exp[-(1/\lambda)S(\traj_j^k)]\delta_j^k}{\sum_{k=1}^{K}\exp[-(1/\lambda)S(\traj_j^k)]}}
\end{equation}
where $\lambda \in \mathbb{R}^+$ is the temperature parameter that weighs the contributions from different control sequences. The first control in the optimal sequence is applied to the system and the entire process is repeated at the next time step.

\begin{mdframed}[style=MyFrame,nobreak=false]
\textbf{Running example \textit{(Safe Planar Navigation)}:} 
As a running example we consider a Dubins' car trying to reach a goal in a cluttered environment. The system can be represented by the following three dimensional dynamics:
\begin{equation}\label{eq:dyn_car} \fontsize{8.5}{10}\selectfont
\dot{\state}
= \begin{bmatrix} \dot{x} & \dot{y} & \dot{\theta} \end{bmatrix}
= \begin{bmatrix} V \cos(\theta) & V \sin(\theta) & \ctrl \end{bmatrix}
\end{equation}
With $x,y$ position of the car's center, $\theta$ its orientation, $V=2 \text{ m/s}$ its fixed speed and $\ctrl \in [-3, 3] \text{ rad/s}$ its angular speed input. The failure set in this case is a $10 \text{m} \times 10 \text{m}$ square enclosure, randomly cluttered with 40 circular obstacles ranging in diameter from 0.35~m to 3.5~m, as shown in Fig ~\ref{fig:mppi_car} (zoomed in) and in Fig.~\ref{fig:dubins_brt} (full scale). The cost function S penalizes the distance to the goal and obstacle penetration (to encourage safety).

Dotted lines in Fig.~\ref{fig:mppi_car} represent possible rollouts of the system following a randomly perturbed control sequence starting from the current state.%We refer to these as the hallucinations of the system for a given state. 

\vspace{1em}
{\centering      \includegraphics[width=0.75\columnwidth]{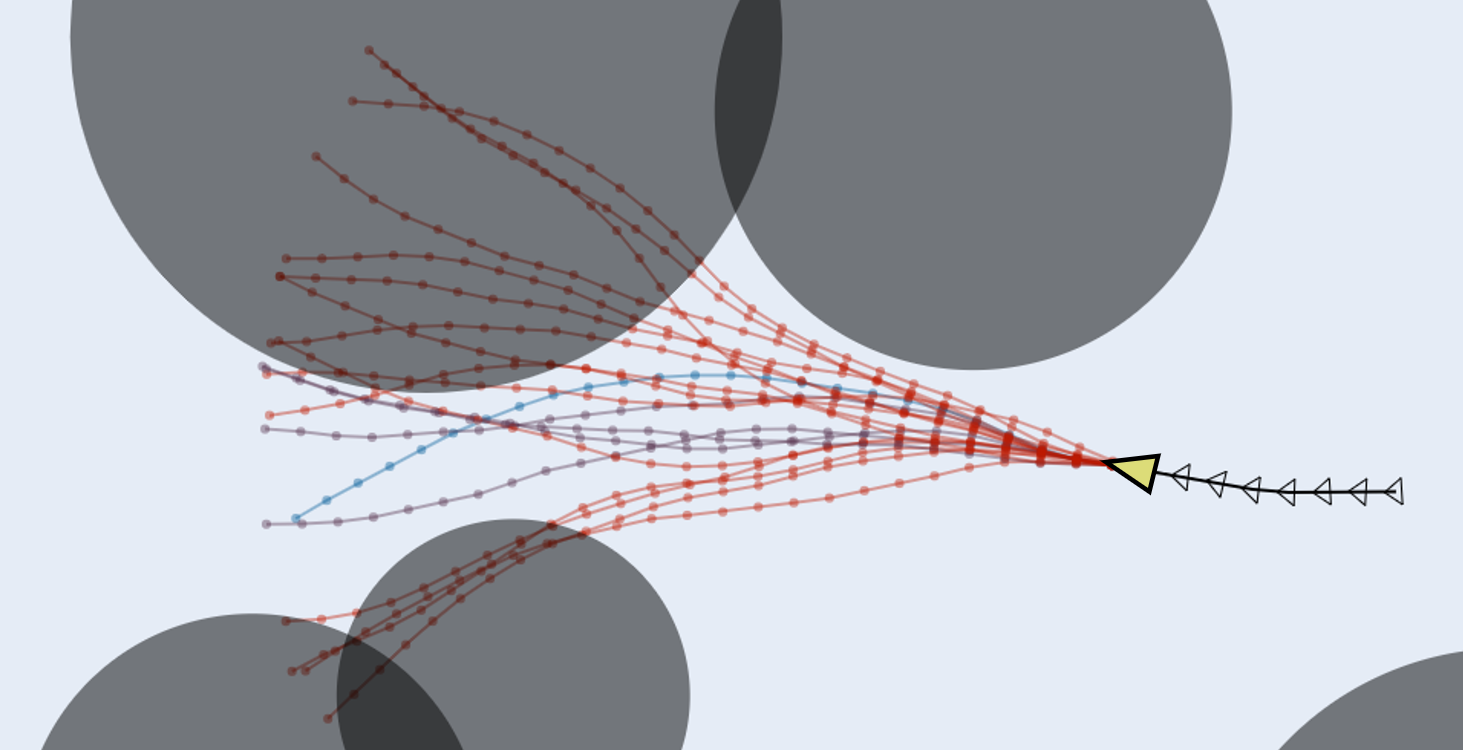}
      \captionof{figure}{Visualization of sampling step in the MPPI algorithm.}
      \label{fig:mppi_car} 
\par}

The rollouts that fail to avoid obstacles achieve a higher cost (shown in red). Thus, the update law in (\ref{eq:update_law}) prioritizes control perturbations that will keep the car safe (shown in a blue-to-purple gradient, with blue indicating lower costs). However, it could be possible that all samples lead to safety violations. Moreover, if the number of safe samples is small, it could lead to a high variance in the performance of MPPI, necessitating the need to account for safety constraints rigorously.

\end{mdframed}

\subsection{\label{background_hj}Hamilton-Jacobi Reachability}

One way to guarantee safe operation of autonomous continuous-time dynamical systems is through Hamilton-Jacobi (HJ) reachability analysis. This approach involves computing the Backward Reachable Tube (BRT) of the failure set $\targetset$. The BRT captures the states from which the system is not able to avoid entering $\targetset$ within some time horizon $\thor$, despite the best control effort.

In HJ reachability, the BRT computation is formulated as a zero-sum game between control and disturbance. Specifically, a cost function is defined to capture the minimum distance to $\targetset$ over time:
\begin{equation}
\costfunctional(\state, \tvar, \cfunc, \dfunc)=\min _{\tdummy \in[\tvar, \thor]} \targetfunc (\state(\tdummy))
\end{equation}
The goal is to capture this minimum distance for optimal system trajectories. Thus, we compute the optimal control that maximizes this distance (drives the system away from the failure set) and the worst-case disturbance signal that minimizes the distance. The value function corresponding to this robust optimal control problem is:
\begin{equation}\label{eq:hji}
 \vfunc(\state, \tvar)=\adjustlimits\inf_{\strat \in \stratset(\tvar)} \sup_{\cfunc} \{\costfunctional(\state, \tvar, \cfunc, \strat[\ctrl](\cdot))\},
\end{equation}
where $\stratset(t)$ defines the set of non-anticipative strategies for the disturbance \cite{bansal2017hamilton}.
The value function in (\ref{eq:hji}) can be computed using dynamic programming, which results in the following final value Hamilton-Jacobi-Isaacs Variational Inequality (HJI-VI) \cite{bansal2017hamilton,lygeros2004reachability,mitchell2005time}:
\begin{equation} \label{eq:pde}\fontsize{9.5}{10}\selectfont
    \begin{aligned}
    \min \{&D_{\tvar} \vfunc(\state, \tvar) + \ham(\state, \tvar, \nabla \vfunc(\state, \tvar)), \targetfunc(\state) - \vfunc(\state, \tvar) \} = 0 \\
    &\vfunc(\state, \horizon) = \targetfunc(\state), \quad \text{for} \ \tvar \in \left[\tinit, \horizon\right]
    % \text{and} \ \state \in \sset
    \end{aligned}
\end{equation}
$D_t$ and $\nabla$ represent the time and spatial gradients of the value function. $\ham$ is the Hamiltonian, which optimizes over the inner product between the spatial gradients of the value function and the dynamics:
\begin{equation}\label{eq:HJIVI_ham_live}
    \ham(\state, \tvar, \nabla \vfunc(\state, \tvar)) = \min_{\ctrl \in \cset} \max_{\dstb \in \dset} \nabla \vfunc(\state, \tvar) \cdot \dyn(\state, \ctrl, \dstb)
\end{equation}
% 
%Intuitively, the term $l(x)-V(x, t)$ in (\ref{eq:pde}) restricts system trajectories that enter and leave the target set, enforcing that any trajectory with a negative distance at any time will continue to have a negative distance for the rest of the time horizon. 
% For a detailed derivation and discussion of the HJI-VI in (\ref{eq:pde}), we refer the interested readers to \cite{mitchell2005time} and \cite{bansal2017hamilton}. 
Once the value function is obtained, the BRT is given as the sub-zero level set of the value function:
\begin{equation}
\brs(\tvar)=\{\state: \vfunc(\state, \tvar) \leq 0\}
\end{equation}
The corresponding optimal safe control can be derived as:
\begin{equation}\label{eq:optctrl}
\ctrl^{*}_{safe}(\state, \tvar)=\argmax_{\ctrl \in \cset} \min_{\dstb \in \dset}\nabla \vfunc(\state, \tvar) \cdot \dyn(\state, \ctrl, \dstb)
\end{equation}
% 
%The system can guarantee reaching the target set as long as it starts inside the BRT and applies the optimal control in (\ref{eq:optctrl}) at the BRT boundary. The optimal adversarial disturbance can be similarly obtained as (\ref{eq:optctrl}).

%This formulation can also be used to provide safety guarantees by switching the roles of the control and disturbance in \eqref{eqn:ham}.
%In that case, the BRT represents the initial states that will eventually be driven into the target by optimal disturbance, despite the best control effort.
%Thus, safety can be maintained as long as the system stays outside the BRT and applies optimal control at the boundary of the BRT.

In safety-ensuring applications, we want to guarantee that the system does not enter $\targetset$ \textit{at any time}. To achieve this, we use a time-converged BRT, as the set of unsafe states typically stops expanding after some time. Since our objective is safety, we will rely exclusively on the converged value function $V(\state)$ throughout this paper. This allows us to synthesize safety controllers using an expression identical to (\ref{eq:optctrl}) but without time dependency.

To compute the value function and obtain the BRT and the optimal controller, we can rely on methods that solve the HJI-VI in \eqref{eq:hji} numerically \cite{mitchell2004toolbox,hj_reach_ASL2023} or using learning-based methods \cite{bansal2021deepreach,fisac2019bridging}.
%bansal2020provably, bui2022optimizeddp
\subsection{\label{background_lr}Least Restrictive Filtering}

If the BRT and associated converged value function are known, a \textit{Least Restrictive Filter} (LRF) can be deployed to guarantee the safe operation of the system.
The LRF is constructed as follows:
\begin{equation}\label{eq:lst_restrict_safety_ctrl}
\controller(\state, \tvar) = \begin{cases}
  \ctrlnom(\state, \tvar) & \vfunc(\state)> 0 \\
   \controller^*_{\text{safe}}(\state) & \vfunc(\state) = 0
\end{cases}
\end{equation}
Here, $\ctrlnom(\state, \tvar)$ corresponds to an arbitrary nominal controller that might optimize a performance criterion without enforcing safety constraints. This control is used when the value function $\vfunc(\state)$ is positive, as the system is not at risk of breaching safety. Whenever the system reaches the boundary of the BRT ($\vfunc(\state)=0$), it switches to $u^{*}_{safe}(x)$ \eqref{eq:optctrl} as this optimal control is guaranteed to maintain or increase $\vfunc(\state)$ keeping the system from entering the unsafe states determined by the BRT, thereby enforcing safety at all times. We refer the reader to \cite{borquezFiltering2023} for details on this filtering technique and proof of the safety guarantees.

\begin{mdframed}[style=MyFrame,nobreak=false]

\textbf{Running example \textit{(Safe Planar Navigation)}:}

We can use the HJI-VI in (\ref{eq:hji}) to calculate the value function and associated BRT, over which we can implement a LRF to guarantee the system's safety. In this case considering the dynamics in \eqref{eq:dyn_car} and the environment definition, we numerically compute the value function using the LevelSetToolbox~\cite{mitchell2004toolbox}.

In Fig.~\ref{fig:mppi_filter}, the purple trajectory represents the system's response to a perturbed control sequence, reaching a critical state at the boundary of the safe set (yellow dot). The boundary of the safe set at this state is illustrated with a dotted line. The red trajectory shows the continuation of the system using the perturbed controls, which leads to the failure set. By contrast, the blue trajectory demonstrates how the unsafe execution can be made safe. This is achieved by applying a LRF step to each control action in the sequence before execution.
% The least restrictive filtering guaranteed safety by using the optimal control given by (\ref{eq:optctrl}) whenever the system's state reaches the boundary of the safe set defined by the BRT, while allowing direct application of the controls if safety is not at risk.

\vspace{0.5em}
{\centering      \includegraphics[width=0.55\columnwidth]{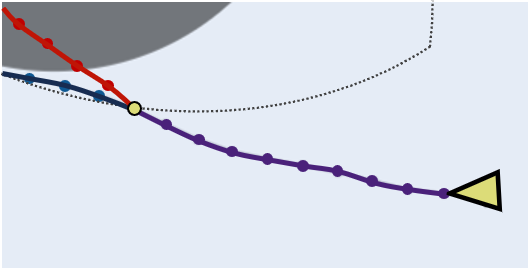}
      \captionof{figure}{Trajectory correction by LRF. System under perturbed controls (purple) reaches the safe set boundary (yellow dot). The red path shows an unsafe execution, while the blue path demonstrates how safety filtering maintains the system outside the failure set.}
      \label{fig:mppi_filter} 
}

\end{mdframed}

%% file: content/04_approach.tex
\section{\label{approach}Dualguard MPPI}

As discussed in Sec. \ref{background_mppi}, MPPI addresses the optimal control problem in \eqref{eq:opt_problem} using a sampling-based method. While MPPI may encourage the satisfaction of safety constraints in \eqref{eq:opt_problem} via penalizing safety violations in the cost function, ensuring safety constraints remains challenging. Moreover, by construction, the high-cost safety breaching sequences are mostly ignored during the optimal control sequence computation, wasting computational resources that could have been used to refine the system performance further. While the safety filtering mechanism discussed in \ref{background_lr} can provide a safety layer after MPPI to enforce the safety constraint, this approach is myopic in nature and might lead to performance impairment in favor of safety.

To overcome these challenges, we propose DualGuard MPPI, a two-layered safety filtering approach. 
Offline, we use HJ reachability to compute a safety value function considering the disturbed system dynamics. By incorporating disturbances at this stage, we ensure safety against worst-case scenarios.
In the first step of our algorithm, safety filtering is incorporated directly into the original MPPI framework, where control sequence rollouts sampled using the undisturbed dynamics are rendered safe by a LRF step.
Disturbances are not included during MPPI propagation to streamline the sampling process, since safety constraints are already handled by the LRF.
This ensures that all rollouts remain within the safe set and can fully contribute to performance optimization.
 To ensure that the resultant optimal control sequence also satisfies the safety constraint, we filter the output optimal control sequence by a safety filter as well. In addition to ensuring the safety constraints, the proposed framework leads to an increased sample efficiency as the safe rollouts keep all sampled trajectories safe and relevant to the performance objective, thereby avoiding “sample wastage” due to safety constraint violation. The proposed algorithm is presented in Alg. 1. Details on the proposed filtering stages are discussed next.

\begin{figure}[t] 
\begin{center} 
\vspace{0.0em}
\includegraphics[width=0.93\columnwidth]{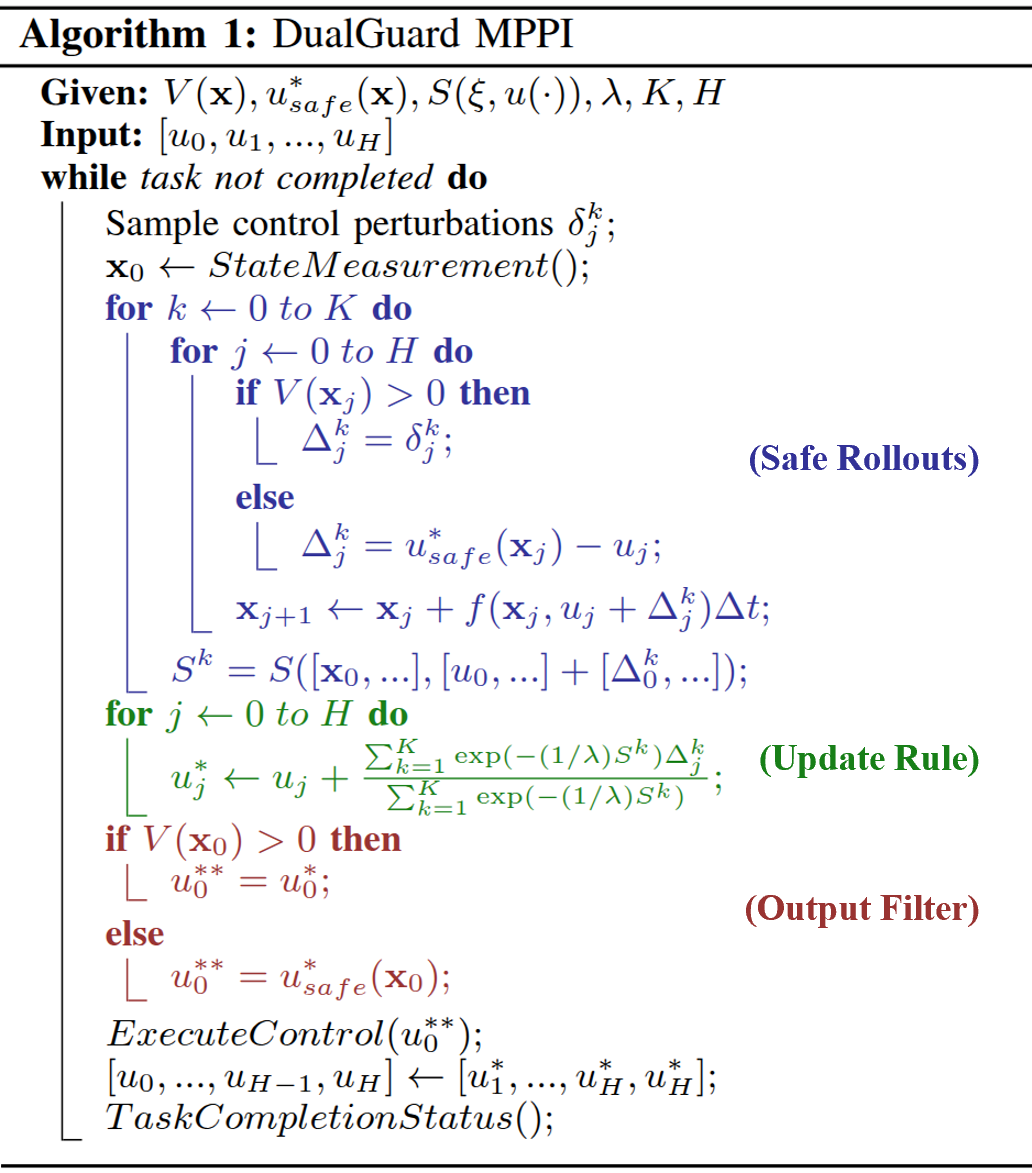}
\vspace{-1em}
\end{center}
\end{figure}
\subsection{\label{safe_hallucinations}Generating Safe Rollouts}
As in classical MPPI, we consider $K$ sequences of random perturbations $\delta_j^k$, that modify a nominal control sequence $u_j$. The perturbed sequences are applied to the deterministic dynamic model of the system while being filtered at each time step along the horizon, using  the LRF in eq. (\ref{eq:lst_restrict_safety_ctrl}), resulting in rollouts that are guaranteed to maintain safety. The cost-to-go $S^k$ for the filtered trajectory is calculated over the safety-filtered control perturbation sequence $\Delta_j^k$.
% obtained as the difference between the filtered control and the nominal sequence.

The nominal control sequence $u_j$, the filtered control perturbations $\Delta_j^k$ and the cost over the safe rollouts $S^k$ are used to calculate the optimal control sequence $u^*_j$ using the update rule defined in (\ref{eq:update_law}). As we weigh over controls that only produce safe trajectories, all K sequences contribute to the performance optimization, reducing the variance of MPPI algorithm and leading to better performance.

\subsection{\label{output_filter}Output Least restrictive filtering}

Even though the optimal control sequence $u^*_j$ was obtained by weighing control perturbations that individually resulted in safe trajectories, the safety guarantees may not hold for $u_j^*$. For this reason, before applying the first control in the sequence $u^*_0$ to the system, we perform one last LRF step to guarantee the safe operation of the complete system. The rest of the sequence is used as the nominal control for the next time step, and the entire process is repeated.

One motivation for this last filtering stage is scenarios where the safe rollouts present multimodality. We can imagine a vehicle trying to swerve around an obstacle by turning right or left; even if both maneuvers result in safe behaviors, a weighting over rollouts split between these modalities could result in maintaining a straight trajectory, causing a collision. Such issues are resolved by this additional filtering stage which picks one of the safe modes.

\begin{mdframed}[style=MyFrame,nobreak=false]

\textbf{Running example \textit{(Safe Planar Navigation)}:}

Considering our running example we observe the proposed steps of DualGuard MPPI in Fig.~\ref{fig:safe_mppi_steps}. First, we visualize the unmodified rollout step in the left panel. Next, using the same control perturbations, we show  how the safe rollout step renders all the samples safe. Sampled trajectories are presented in a blue-to-red scale, corresponding to the associated low-to-high costs.

The right panel of Fig.~\ref{fig:safe_mppi_steps} shows how the safe rollouts might present themselves in a multimodal fashion for some obstacle configurations; this indicates a possible failure mode where the control sequence given by the update rule in \eqref{eq:update_law} could drive the vehicle directly into the obstacles, leading to safety violations. The proposed output filtering stage safeguards against this possibility.  

\vspace{1em}
{\centering      \includegraphics[width=1.0\columnwidth]{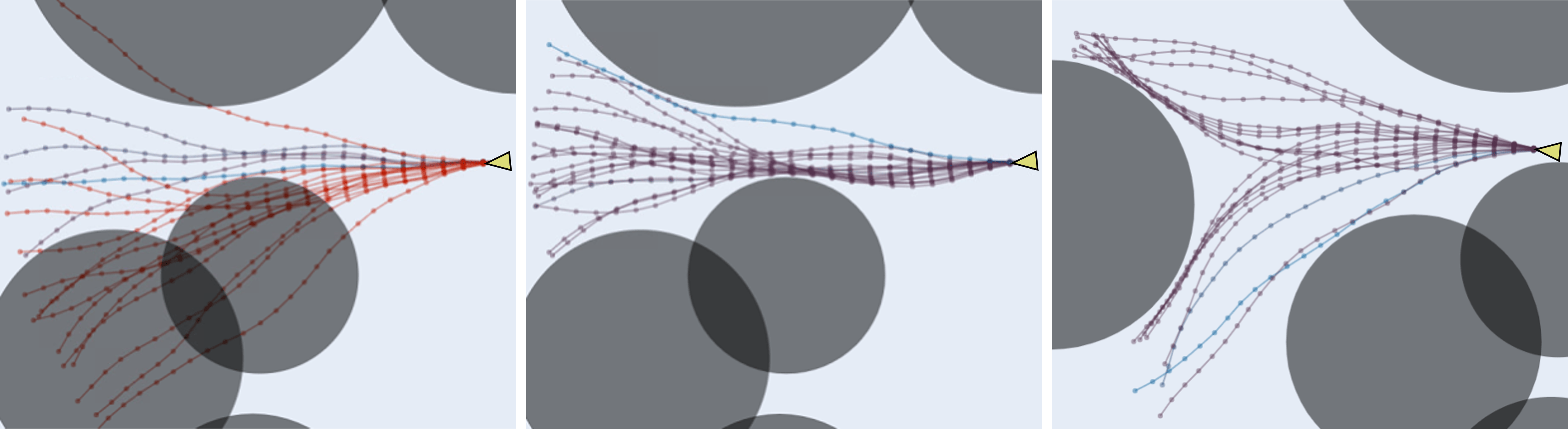}
      \captionof{figure}{Unmodified rollouts (Left). Safe rollouts (Center). Possible multimodality on safe rollouts encourages the use of output least restrictive filtering (Right).}
      \label{fig:safe_mppi_steps}\par} 

\end{mdframed}

% %
% \definecolor{darkblue}{rgb}{0.2, 0.2, 0.6}
% \definecolor{darkgreen}{rgb}{0.1, 0.5, 0.1}
% \definecolor{darkred}{rgb}{0.6, 0.2, 0.2}
% \LinesNotNumbered
% \begin{algorithm}
%   \caption{DualGuard MPPI}\label{alg1} 
% \small % Adjust font size to \small; try \footnotesize or \scriptsize for even smaller sizes
% \textbf{Given:} $\vfunc(\state), \ctrl^*_{safe}(\state),S(\traj,\ctrl(\cdot)),\lambda, K, H$\\
%  \textbf{Input:} $[u_0,u_1,...,u_H]$\\
%  % \textbf{Output:} $[u_1^*,...,u_H^*],u_0^{**}$\\
% \While {task not completed}{
% Sample control perturbations $\delta_j^k;$\\
% $\state_0 \gets StateMeasurement();$
% \textcolor{darkblue}{\\
% \textbf{(Safe Hallucinations)}\\ % Section label
% \For{$k\gets 0\; to\; K$}{ 
%     \For{$j\gets 0\; to\; H$}{
%          \If{$\vfunc(\state_j)>0$}{$\Delta_j^k = \delta_j^k;$}
%          \Else{$\Delta_j^k = \ctrl^*_{safe}(\state_j)-\ctrl_j;$}
%          $\state_{j+1} \gets \state_j + \dyn(\state_j, \ctrl_j + \Delta_j^k)\Delta\tvar;$\\
%      }%for end
%      $S^k = S([\state_0,...],[u_0,...]+[\Delta_0^k,...]);$
% }}%for end
% \textcolor{darkgreen}{\\
% \textbf{(Update Rule)}\\ % Section label
% \For{$j\gets 0\; to\; H$}{
% $u_j^{*}\gets u_j+{\frac{\sum_{k=1}^{K}\exp(-(1/\lambda)S^k)\Delta_j^k}{\sum_{k=1}^{K}\exp(-(1/\lambda)S^k)}};$
% }}
% \textcolor{darkred}{\\
% \textbf{(Output Filter)}\\
% \If{$\vfunc(\state_0) > 0$}{$u_0^{**} = u_0^{*};$}
% \Else{$u_0^{**} = \ctrl^*_{safe}(\state_0);$}
% }
% $ExecuteControl(u_0^{**});$\\
% $[u_0,...,u_{H-1},u_H] \gets [u_1^*,...,u_H^*,u_H^*];$\\
% $TaskCompletionStatus();$\\
% }%while end
% \end{algorithm}

%% file: content/05_results.tex
\section{\label{cases}Simulation Studies}

We evaluate the performance of our proposed method on simulation studies and a real hardware testbed. For this, we compare the proposed approach against five baseline methods. Each baseline builds on the MPPI control approach but incorporates distinct modifications to enhance safety.\par

\noindent \textit{Obstacle Penalty (Obs. penalty)}: A classical MPPI approach to collision avoidance by adding a high penalty for entering the obstacles in the cost function, incentivizing the system to avoid these penalized regions.

\noindent \textit{BRT Penalty (BRT penalty)}: Advanced MPPI methods penalize sampled trajectories that enter a safe zone defined as augmented obstacles, Discrete Barrier States (DBaS), or Control Barrier Functions (CBF) \cite{discrete_barrier_2024,parwana2024modelpredictivepathintegral,RAPA_2024}. To represent this technique, we implement a baseline that penalizes trajectories that enter the obstacle set BRT. This preemptive approach penalizes trajectories that have entered states from which collision is unavoidable, even if they have not yet entered the obstacle set within the sampling horizon.

\noindent \textit{Obstacle Penalty with Output LRF (Obs. pen. + LRF)}: Extension of the Obstacle Penalty baseline by adding a LRF step. This corresponds to a naive safety filtering approach where the LRF is used to correct any control that would result in a safety violation just before it is applied.

\noindent \textit{BRT Penalty with Output LRF (BRT pen. + LRF)}: Similar to the previous baseline, this extends the BRT Penalty baseline by adding least restrictive filtering safety guarantees.

\noindent \textit{Shield Model Predictive Path Integral (Shield MPPI)}: We implement the approach described in \cite{mppi_shield}, where a cost related to the CBF condition is considered for rollouts, followed by an approximate CBF-based repair step over the updated control to promote safety. To ensure a fair comparison, we use the HJ value function as a barrier function so that all methods work on the maximal safe set \cite{cbvf}.

\noindent \textbf{Evaluation Metrics:} To quantitatively assess the performance of each method, we define the following evaluation metrics. These metrics will be measured over batches of simulation runs for each method.

\begin{itemize}
    \item Failure: Percentage of episodes that result in a safety violation (e.g., a collision).
    \item RelCost: Average cost of the trajectories and standard error, normalized to our method's mean cost. We only consider common safe trajectories between the compared method and ours. This allows for a fair comparison and ensures that the RelCost is not affected by the high failure penalties for some of the baselines.
    % \item CompTime: Average computation time required to perform each iteration of the controller.
\end{itemize}

Furthermore, when deemed relevant, we define additional performance-related metrics specific to each case study.

\subsection{\label{case_dub3D}Safe Planar Navigation}

We assess the running-example safe planar navigation with dynamics defined in \eqref{eq:dyn_car}. In each episode, the system must navigate from an initial state $\state_0$ to a goal region of radius 0.1~m in the $xy$ plane centered at the goal state $\state_g$, within a time horizon of $T = 20\text{ s}$. The cost function is defined as the squared distance from the goal, a control effort penalty, and a method dependent safety penalty $P(\state)$:
\begin{equation}\fontsize{9.5}{10}\selectfont
S=\sum_{t=0}^{T} (\state_t - \state_g)^T Q(\state_t - \state_g) + m(u_t)^2 + P(\state_t) \\
\end{equation}
%
% \noindent with $Q = \text{diag}([1, 1, 0])$, $m = 0.2$, and $P=10000$ if the state is inside the obstacles or the BRT (corresponding to that controller's cost penalty type) and $K=0$ otherwise.

We test each controller in the same set of 100 episodes defined by pairs $(\state_0, \state_g)$. These initial-goal state pairs were selected randomly but constrained to be outside the BRT, near the environment boundary, and separated by at least 5 m for diversity.
The results are summarized in Table~\ref{tab:dubins_results}. In addition to the metrics previously defined, we also compute the Success rate, defined as the percentage of episodes that reach the goal region safely within the time horizon, and Timeout, defined as the percentage of episodes that fail to reach the target within the given time horizon. 
% Computations for this experiment were performed on CPU using an AMD Ryzen 5 3600X.

\begin{figure}[t] 
\begin{center} 
\vspace{0.0em}
\includegraphics[width=0.99\columnwidth]{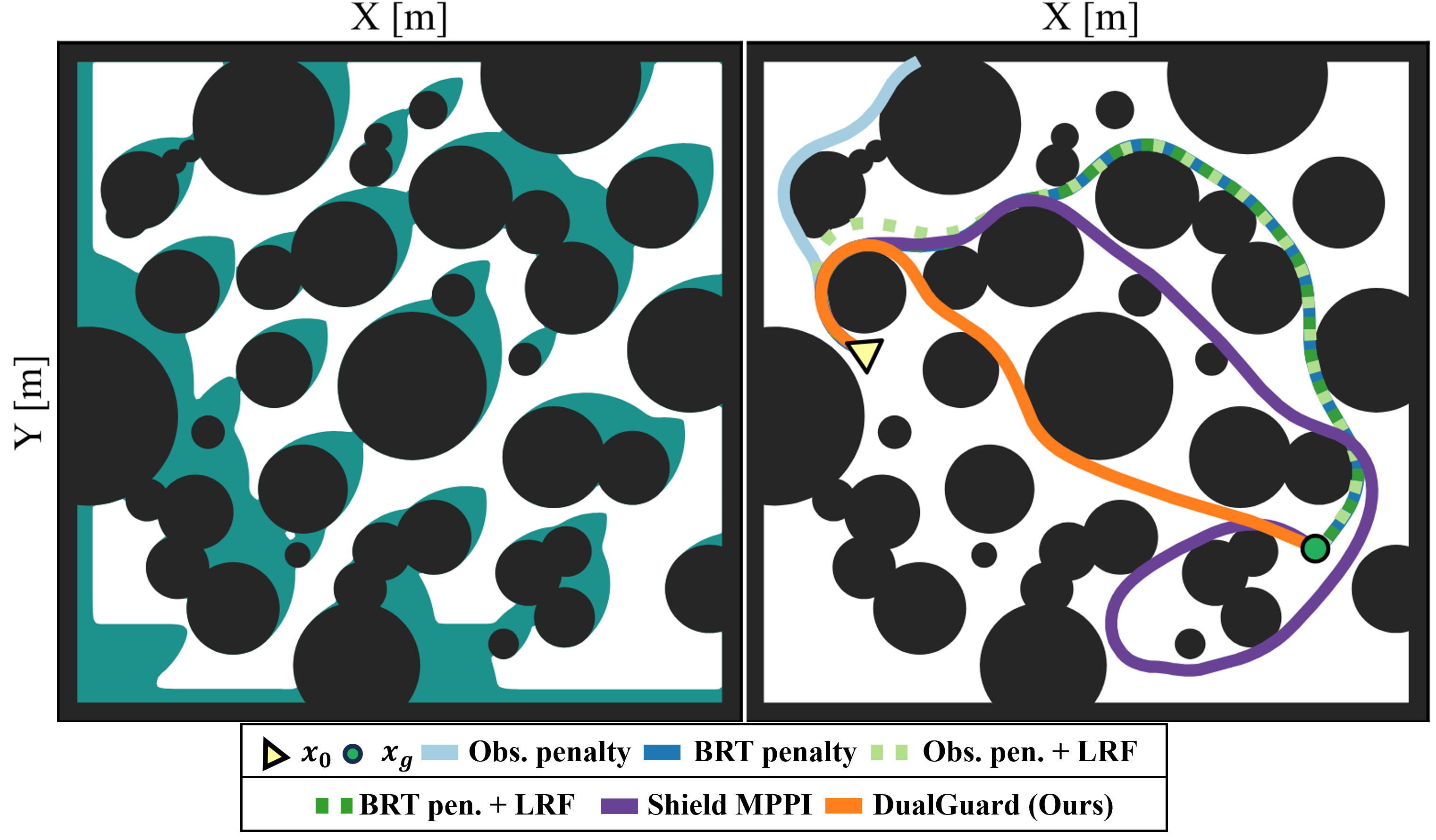}
\vspace{-1.5em}
\caption{Safe Planar Navigation environment, with obstacles in gray and a slice of the BRT for a southwestern heading is in teal.}
\vspace{-1.2em}
\label{fig:dubins_brt}
\end{center}
\end{figure}

\setlength\tabcolsep{1pt}
\begin{table}[b]
\fontsize{7pt}{7pt}\selectfont
\caption{Safe Planar Navigation results over 100 episodes. (*) Indicates statistically significant differences ($p < 0.05$) compared to our method.}
\centering
\renewcommand{\arraystretch}{1.2}   % Adjust row height for vertical centering
\begin{tabularx}{\columnwidth}{|>{\centering\arraybackslash}p{0.8cm}|>{\centering\arraybackslash}p{2.2cm}|>{\centering\arraybackslash}p{1.0cm}|>{\centering\arraybackslash}p{1.0cm}|>{\centering\arraybackslash}p{1.0cm}|>{\centering\arraybackslash}X|}
\hline
\textbf{K}&\textbf{Method}&\textbf{Success}&\textbf{Timeout}&\textbf{Failure}&\textbf{RelCost} \\
\hline
\multirow{6}{*}{1000}
& Obs. penalty    & 49 &  1 & 50 & {1.19 ± 0.14*} \\
& BRT penalty     & 72 & 21 &  7 & {2.05 ± 0.17*} \\
& Obs. pen. + LRF & 80 & 20 &  \textbf{0} & {1.44 ± 0.10*} \\
& BRT pen. + LRF  & 74 & 26 &  \textbf{0} & {2.61 ± 0.29*} \\
& Shield MPPI     & 83 & 17 &  \textbf{0} & {1.70 ± 0.14*} \\
& DualGuard (Ours)      & 99 &  1 &  \textbf{0} & {1.00 ± 0.07}  \\
\hline
\multirow{6}{*}{250}
& Obs. penalty    & 31 &  0 & 69 & {0.99 ± 0.10} \\
& BRT penalty     & 58 & 15 & 27 & {2.28 ± 0.23*} \\
& Obs. pen. + LRF & 81 & 19 &  \textbf{0} & {1.80 ± 0.14*} \\
& BRT pen. + LRF  & 69 & 31 &  \textbf{0} & {5.46 ± 0.59*} \\
& Shield MPPI     & 78 & 21 &  1 & {1.85 ± 0.14*} \\
& DualGuard (Ours)      & 98 &  2 &  \textbf{0} & {1.00 ± 0.08} \\
\hline
\multirow{6}{*}{60}
& Obs. penalty    & 11 &  0 & 89 & {1.07 ± 0.16} \\
& BRT penalty     & 43 & 21 & 36 & {3.70 ± 0.49*} \\
& Obs. pen. + LRF & 70 & 30 &  \textbf{0} & {2.45 ± 0.18*} \\
& BRT pen. + LRF  & 55 & 45 &  \textbf{0} & {6.48 ± 0.60*} \\
& Shield MPPI     & 64 & 36 &  \textbf{0} & {2.82 ± 0.20*} \\
& DualGuard (Ours)      & 96 &  4 &  \textbf{0} & {1.00 ± 0.07} \\
\hline
\end{tabularx}

\label{tab:dubins_results}
\end{table}

Examining the results for $K=1000$, we see that DualGuard achieves zero failure rate and highest success rate. While all methods with an output LRF avoid collisions as expected, our method also times-out in fewer episodes, indicating a better performance. The safety-performance co-optimization capabilities of the proposed method are also evident from its lowest RelCost across the safe methods, this difference was further validated to be statistically significant through paired t-tests against DualGuard-MPPI.

% all for 1000 samples in Episode 23
To understand the different controllers' behavior, consider the trajectories shown in Figure~\ref{fig:dubins_brt}. Here, the obstacle penalty method fails early by entering a region in the northeast corner which is impossible to escape and hence avoided by all BRT-aware methods. The other baseline methods execute a reasonable trajectory to the goal, but the proposed method manages to find a lower-cost trajectory by taking a shortcut after rounding the first obstacle. This is because the baseline methods are still fundamentally limited by the exploration capabilities of MPPI -- if the underlying samples are unsafe, they contribute little to the performance optimization. On the other hand, the proposed method generates safe rollouts, allowing it to explore more relevant parts of the environment with the same number of samples, as previously visualized in Fig.~\ref{fig:safe_mppi_steps}, leading to a better performance optimization.
% Thus the safe-sampling process of the proposed method improves the exploratory properties of the algorithm without any additional exploration hyperparameter.

A consequence of the superior exploration capabilities of the proposed method is that it can achieve similar success rates with far fewer samples. We summarize the results for different methods as we vary the number of samples in Table~\ref{tab:dubins_results}. We note that while all methods degrade with fewer samples as expected, DualGuard remains relatively consistent. With only 60 samples, it succeeds in 96\% of the episodes, with the closest contender not going over 70\%.
This could enable its deployment on resource-constrained systems, which MPPI methods typically struggle with.% ( indicated by the poor performance of MPPI baselines for fewer samples).

\subsection{\label{case_drone6D}Autonomous Quadrotor Navigation}
Next, we demonstrate our method on a 6-dimensional nonlinear quadrotor system. 
The system follows the dynamics:
%
% \setlength{\arraycolsep}{2pt} % Adjust spacing as needed
% \begin{equation}\label{eq:dyn_hw}
% {\fontsize{8.5}{10}\selectfont
% \dot{\state}
% = \begin{bmatrix} \dot{x} & \dot{y} & \dot{\theta} \end{bmatrix}
% = \begin{bmatrix} V \cos(\theta)+ d_x, & V \sin(\theta)+ d_y, & V \tan(\delta) / L \end{bmatrix}
% }
% \end{equation}
% \setlength{\arraycolsep}{5pt} % Reset spacing
% \vspace{-1.5em}
\begin{equation} \fontsize{8.5}{10}\selectfont
\begin{aligned}
    \dot{\state} &= \begin{bmatrix} \dot{x} & \dot{y} & \dot{z} & \dot{v_x} & \dot{v_y} & \dot{v_z} \end{bmatrix}^T \\
    &= \begin{bmatrix} v_x & v_y & v_z & a_g \tan u_\theta & -a_g \tan u_\phi & u_T - a_g \end{bmatrix}^T
\end{aligned}
\end{equation}
% good enouth for now

% \begin{equation}
% \dot{\state}
% = \begin{bmatrix} \dot{x} \\ \dot{y} \\ \dot{z} \\ \dot{v_x} \\ \dot{v_y} \\ \dot{v_z} \end{bmatrix}
% = \begin{bmatrix} v_x \\ v_y \\ v_z \\  a_g \tan u_\theta \\ -a_g \tan u_\phi \\ u_T - a_g \end{bmatrix}
% \end{equation}

where $a_g = 9.81~\text{m/s}^2$ is gravity, and the controls $u = [u_T, u_\phi, u_\theta]$ correspond to thrust, roll, and pitch angles. Their ranges are $u_T \in [7.81, 11.81]~\text{m/s}^2$ and $u_\phi, u_\theta \in [-17.18^\circ, 17.18^\circ]$. Zero-mean Gaussian noise is added to simulate actuation errors.

%where $a_g = 9.81 \text{ m/s}^2$ is the gravity acceleration, and acceleration $u_T$, roll $u_\phi$, and pitch $u_\theta$ are the controls $u = [u_T, u_\phi, u_\theta]$ of the system with ranges $u_T \in [7.81, 11.81] \text{ m/s}^2$ and $u_\phi, u_\theta \in [-17.18^{\circ},17.18^{\circ}]$. Zero mean Gaussian noise is added to the actions to simulate real-world actuation errors.

\begin{figure}[h]
\begin{center}
\vspace{0.0em}
\includegraphics[width=.95\columnwidth]{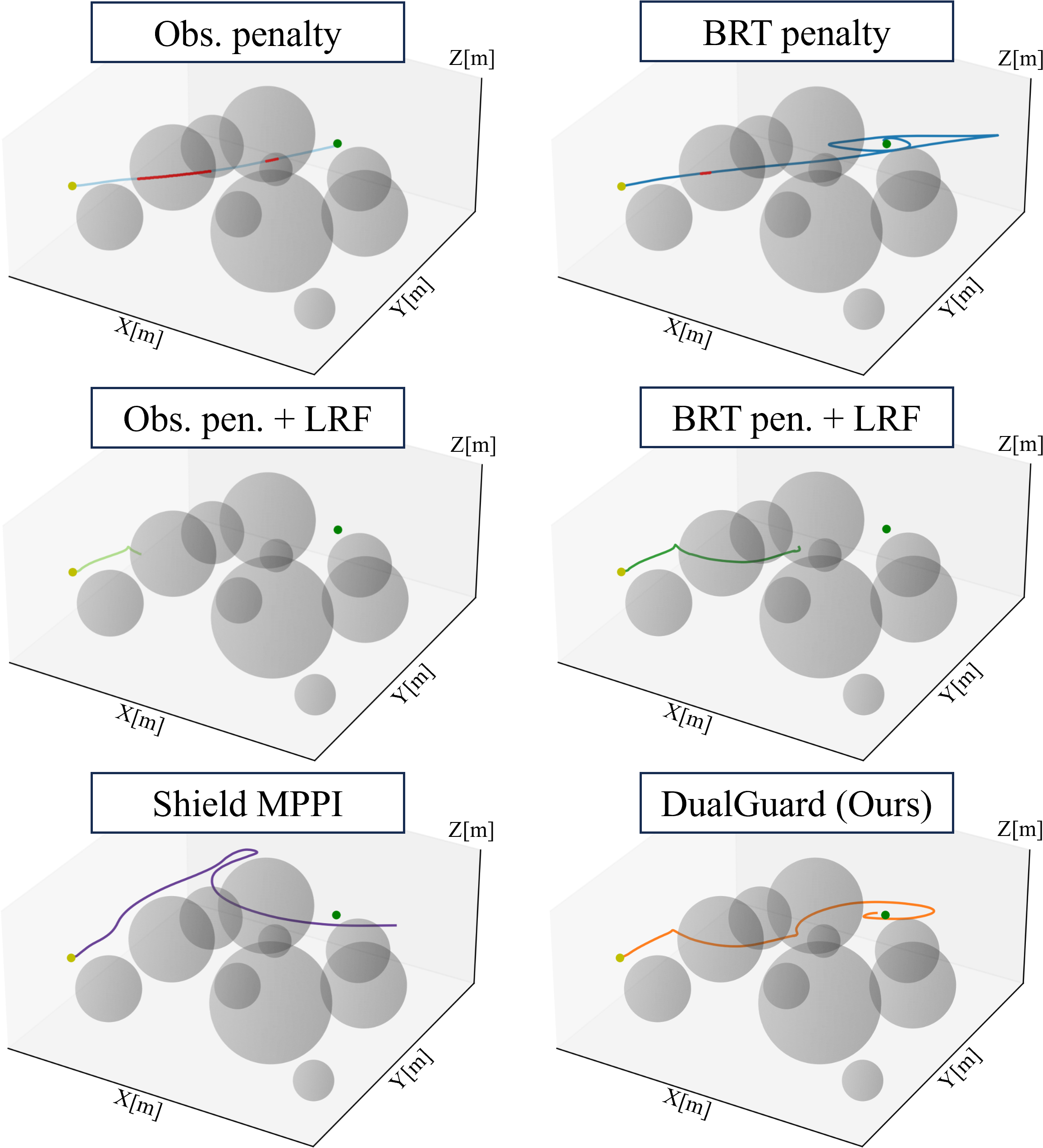}
% \vspace{-0.8em}
% \caption{Left: The quadrotor environment with obstacles and goal location. Trajectories of the quadrotor for each method are shown in different colors. Right: The hallucinations for Obs cost, BRT cost, and the proposed method.\ucnote{the figure is a placeholder}}
\caption{Quadrotor environment with obstacles (gray), initial position (olive), and goal (dark green). Trajectories for each method are shown in different colors, obstacle intrusions are denoted by red.}
\vspace{0em}
\label{fig:quadrotor}
\end{center}
\end{figure}

The environment is a $10 \text{m} \times 10 \text{m} \times 4 \text{m}$ room containing 10 spherical obstacles of varying sizes.
$V(\state)$ and the associated BRT of the system are numerically computed using the LevelSetToolbox \cite{mitchell2004toolbox}.
The task of the quadrotor is to reach a goal sphere with radius $0.3m$ around the goal state $\state_g$ (shown by the green sphere in Fig.~\ref{fig:quadrotor}), starting from a random initial position $\state_0$ while avoiding collisions with the obstacles.
The system must reach the goal location within a time horizon of $T = 10\text{s}$; otherwise, the episode is considered timeout.
If the system collides with an obstacle or the environment boundary, the episode is considered unsafe.

The cost function consists of distance to the goal, control effort penalty, and a method dependent safety penalty $P(\state)$:

\begin{equation}
\small % or \footnotesize or \scriptsize for different sizes
    \begin{aligned}\label{eq:cost_mppi_quadrotor}
    % {\mathcal{L}}(\state(\tvar),\ctrl(\tvar),\tvar)
    S = & (\state_t - \state_g)^T Q(\state_t - \state_g) \\
    &+ \| (1/\ctrl_{range})^T (\ctrl_t - \ctrl_{mean})\|_2^2 + P(\state_t)
    \end{aligned}
\end{equation}

where
$Q = \text{diag}([20, 20, 20, 0, 0, 0])$,
$\ctrl_{range} = [2, 0.3, 0.3]$ and $\ctrl_{mean} = [9.81, 0, 0]$. For all methods, $500$ rollouts are simulated for a horizon of $0.3s$. Each method is tested starting from the same 100 randomly sampled initial positions over the environment with at least $5m$ distance from the goal location. The results are summarized in Table~\ref{tab:quadrotor_results}.
%
% $K_{Obs}$ and $K_{BRT}$ are the penalty weights for entering the obstacle set and BRT respectively.
% The penalty weights $K_{Obs}$ and $K_{BRT}$ alternate between $(K_{Obs},K_{BRT})=(10,0)$ for the 'Obs. penalty' and 'Obs. pen. + LRF' cases and $(K_{Obs},K_{BRT})=(0,10)$  for the 'BRT penalty' and 'BRT penalty + LRF' cases.
% where the safety rate is defined as the percentage of episodes where the system doesn't collide.
% The success rate is defined as the percentage of episodes where the system reaches the goal location without colliding.
% To compare the performance of the controllers, we compute the accumulated cost of the successful trajectories common to all methods using $K_{Obs}=10$ and $K_{BRT}=0$.

In this case, the obstacle penalty method leads to a poor success rate. It is partially due to the added disturbance in the system that makes it challenging to ensure safety. A BRT penalty improves the safety rate from 14\% to 30\%, but it still does not enforce safety, as shown in Fig.~\ref{fig:quadrotor}. 
By penalizing decreases in safety and the approximate repair step , Shield MPPI decreases the number of collisions to 21.
In contrast, safety is guaranteed for baselines with a LRF. Nevertheless, a significant portion of the rollouts end up inside obstacles and are unable to contribute to the control synthesis meaningfully. Therefore, methods with safety-filtered actions experience many trajectories that fail to reach the goal within the given task horizon despite maintaining safety; in Fig.~\ref{fig:quadrotor}, we show such timeout executions. 
By filtering the rollouts, our method incorporates samples that would otherwise be unmeaningful into the optimization. Doing so achieves the best success and safety rates among all methods and achieves better cost on safe trajectories compared to the others that guarantee safety. 
\begin{table}[h]
\caption{Quadrotor simulation results. Asterisks (*) indicate statistically significant differences ($p < 0.05$) compared to our method.}
% Standart error is provided
\centering
\begin{tabularx}{\columnwidth}{|>{\centering\arraybackslash}X|>{\centering\arraybackslash}p{1.2cm}|>{\centering\arraybackslash}p{1.2cm}|>{\centering\arraybackslash}p{1.2cm}|>{\centering\arraybackslash}p{2.3cm}|}
\hline
\textbf{Method} & \textbf{Success} & \textbf{Timeout} & \textbf{Failure} & \textbf{RelCost} \\ 
\hline
Obs. penalty & 14 & 0 & 86             & {0.48 $\pm$ 0.06*} \\ 
BRT penalty & 30 & 0 & 70             & {0.60 $\pm$ 0.09*} \\ 
Obs. pen. + LRF & 68 & 32 & \textbf{0}   & {1.22 $\pm$ 0.06*} \\ 
BRT pen. + LRF & 65 & 35 & \textbf{0}   & {1.21 $\pm$ 0.06*} \\ 
Shield MPPI & 69 & 10 & 21             & {0.91 $\pm$ 0.05} \\ 
DualGuard (Ours) & \textbf{75} & 25 & \textbf{0} & {1.00 $\pm$ 0.05} \\
\hline
\end{tabularx}
\label{tab:quadrotor_results}
\end{table}
%\vspace{-1em}

%% file: content/06_hardware.tex
\section{\label{case_rc_car}Hardware Experiments - RC Car}

Finally, we consider a real-world miniature RC car with dynamics modeled as \eqref{eq:dyn_hw}, with $L=23.5 \text{ cm}$, controls $\ctrl=[V,\delta]$ with ranges $V \in [0.7,1.4]\text{ m/s}$ and $\delta \in [-25^{\circ},25^{\circ}]$, and disturbances $d_x,d_y \in [-0.1,0.1]$ to account for model mismatches and state estimation error. The vehicle's task is to complete laps on the racetrack shown in Fig.~\ref{fig:exp_main_result}.
\setlength{\arraycolsep}{2pt} % Adjust spacing as needed
\begin{equation}\label{eq:dyn_hw}
{\fontsize{8.5}{10}\selectfont
\dot{\state}
= \begin{bmatrix} \dot{x} & \dot{y} & \dot{\theta} \end{bmatrix}
= \begin{bmatrix} V \cos(\theta)+ d_x, & V \sin(\theta)+ d_y, & V \tan(\delta) / L \end{bmatrix}
}
\end{equation}
\setlength{\arraycolsep}{5pt} % Reset spacing

We adapt our evaluation metrics to better reflect real-world, single-run applicability. Instead of batch statistics, we measure the \textit{CompTime} for each method -- the time taken to generate and evaluate potential samples within a multiple-lap run. This metric reflects how well-suited each technique is for real-time control. We also report the car's average \textit{Speed} over three laps to measure how aggressive the policy is. The \textit{RelCost} metric remains consistent with the simulations.

\begin{figure}[h!] 
\begin{center} 
\vspace{0.0em}
\includegraphics[width=0.99\columnwidth]{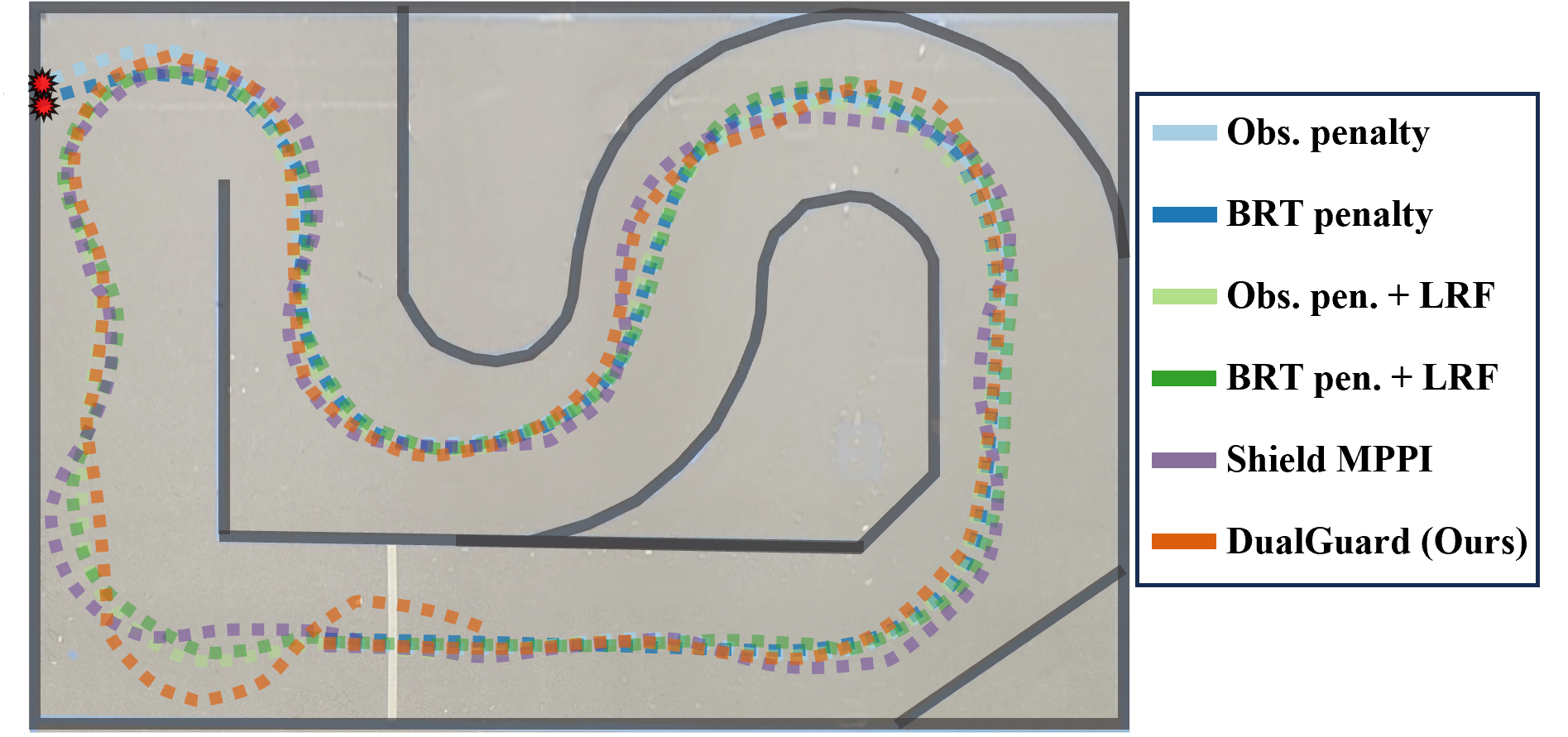}
\vspace{0.0em}
\caption{Top view of the RC car's trajectories under each method.}
\label{fig:hw_traj}
\end{center}
\end{figure}

As cost function we use (\ref{eq:cost_mppi_car}), where the first term penalizes going slower than $V_\text{max}=1.4\text{ m/s}$, the second term penalizes the distance from the track's center line, the third term $P(\state)$ is a method dependent safety penalty. The value function and associated BRT are numerically computed using \cite{mitchell2004toolbox}.
%, and the last term penalizes going into the BRT. The penalty weights $K_{Obs}$ and $K_{BRT}$ alternate between $(K_{Obs},K_{BRT})=(50,0)$ for the 'Obs costs' and 'Obs costs+ LR filter' cases and $(K_{Obs},K_{BRT})=(0,50)$  for the 'BRT costs' and 'BRT costs+ LR filter' cases, for the proposed method their value is irrelevant as hallucinations are guaranteed not to enter the BRT or obstacle sets. 
%with $l_{center}$ the constant distance between the center of the lane and the nearest edge,
\begin{equation}\small
\label{eq:cost_mppi_car}
S = (V_{max}-V)^2 + K_{c} (l_{center}-\targetfunc(x)) +  P(\state)
\end{equation}
The controllers were implemented using JAX \cite{jax} on a laptop equipped with an NVIDIA GeForce RTX 4060. 
% for its parallelization capabilities; the GPU used is a Laptop NVIDIA GeForce RTX 4060 where 
We generate $1000$ parallel rollouts (with $100$ time steps each) in a loop running at $50Hz$. Results are summarized in Table~\ref{tab:hw_results}, and trajectories for the first lap are shown in Fig~\ref{fig:hw_traj}.

First, we highlight the need for hard safety constraints as methods that only rely on safety penalties fail to clear the top-left tight turn as shown in Fig~\ref{fig:hw_traj}. Fine-tuning the cost function and MPPI parameters might allow unfiltered methods to complete laps. Still, we want to consider and compare methods that provably allow for safe executions.

\begin{table}[t]
\caption{Hardware experiments results summary.}
\centering
\renewcommand{\arraystretch}{1.2} % Adjust row height for vertical centering
\begin{tabularx}{\columnwidth}{|>{\centering\arraybackslash}p{2.5cm}|>{\centering\arraybackslash}X|>{\centering\arraybackslash}p{1.4cm}|>{\centering\arraybackslash}p{2.1cm}|}
\hline
\textbf{Method} & \textbf{CompTime (ms)} & \textbf{RelCost} & \textbf{Speed (m/s)} \\
\hline
Obs costs & 1.8 ± 0.3 & fail & 1.00 ± 0.05 \\
BRT costs & 1.8 ± 0.3  & fail & 1.01 ± 0.06 \\
Obs costs + LRF & 1.7 ± 0.4 & 1.1874 & 1.03 ± 0.12\\
BRT costs + LRF & 1.8 ± 0.4 & 1.1626 & 1.04 ± 0.12 \\
Shield-MPPI & 1.7 ± 0.2 & 1.1038 & 1.04 ± 0.08\\
DualGuard (Ours) & 2.5 ± 0.4 & 1.0000 & 1.10 ± 0.11 \\
\hline
\end{tabularx}
\label{tab:hw_results}
\end{table}
DualGuard leads to faster and more performant trajectories than the other safe baselines. Comparison with the LRF baselines shows how the proposed safe rollout step improved the quality of the samples as seen in Fig.~\ref{fig:exp_main_result}(B)(C), leading to better overall performance and higher average speed. Also, the proposed method outperforms Shield-MPPI even after tuning its hyperparameters to the best of our capabilities to maintain safety without an excessive impact on performance.

Computation times are nearly identical across baselines, as each method fundamentally involves calculating performant terms of the cost function and querying the obstacle set or BRT for safety-related penalties. DualGuard introduces an additional LRF step for each sample along the trajectories, resulting in an increase in computational time. Nevertheless, all methods operate well within the $20ms$ time budget, leaving ample time for the control loop to handle state estimation, communications, and actuation.

%% file: content/07_conclusion.tex
\section{Limitations and Future Work}
In this work, we have presented DualGuard-MPPI, a novel MPPI framework for addressing the safety-constrained optimal control problem. By integrating Hamilton-Jacobi reachability analysis with MPPI, our approach ensures strict adherence to safety constraints throughout both the sampling and control execution phases. This combination enables high-performance trajectory optimization without compromising safety, validated through extensive simulation and real-world experiments. DualGuard-MPPI stands out for its capability to eliminate safety-related terms from the cost function, thereby streamlining the optimization process to focus purely on performance objectives.

While these contributions establish DualGuard-MPPI as a robust and scalable framework, certain limitations remain that warrant further exploration.
The requirement for a precomputed BRT introduces considerable computational overhead during setup, which may hinder its implementability in rapidly changing environments.
% \hl{Furthermore, incorporating disturbance-aware MPPI frameworks could enhance performance by allowing the controller to proactively account for disturbances during trajectory optimization}~\cite{RA_MPPI}.
Additionally, the reliance on explicitly defined system dynamics may restrict the framework's applicability to systems with highly complex or partially unknown models. Addressing these challenges—such as by leveraging online reachability methods to dynamically update BRTs~\cite{warmStart,borquez2023param_cond_reach} or using reachability methods for black-box systems~\cite{blackboxreach2024,RL_reach2021} could significantly expand the framework's usability. In addition, we will explore the deployment of the proposed approach on other safety-critical robotics applications.

% Our key contributions include the development of a "safe hallucination" step, guaranteeing safe and efficient sampling. This innovation improves sampling relevance, reduces computational waste, and provides deterministic safety guarantees. Furthermore, the incorporation of a final output filter enhances robustness by safeguarding against potential failures in combining individually safe samples. These contributions position DualGuard-MPPI as a versatile and scalable solution for safety-critical robotic applications.